\documentclass[letter,longauth]{aa}

\usepackage[english]{babel}
\usepackage{graphicx}
\usepackage{setspace}
\usepackage[normalem]{ulem}
\usepackage{units}
\usepackage[euler]{textgreek}
\usepackage{hhline}
\usepackage{natbib}
\bibpunct{(}{)}{;}{a}{}{,} 
\usepackage[modulo, switch]{lineno}

\def\Fermi{\textit{Fermi}-LAT\,}

\begin{document}

%
%
\title{Detection of the Geminga pulsar with MAGIC hints at a power-law tail emission beyond 15 GeV}
\titlerunning{Detection of the Geminga pulsar with MAGIC} 

%
%
\author{
V.~A.~Acciari\inst{1} \and
S.~Ansoldi\inst{2} \and
L.~A.~Antonelli\inst{3} \and
A.~Arbet Engels\inst{4} \and
K.~Asano\inst{5} \and
D.~Baack\inst{6} \and
A.~Babi\'c\inst{7} \and
A.~Baquero\inst{8} \and
U.~Barres de Almeida\inst{9} \and
J.~A.~Barrio\inst{8}
% ...
}
\institute { Inst. de Astrof\'isica de Canarias, E-38200 La Laguna, and Universidad de La Laguna, Dpto. Astrof\'isica, E-38206 La Laguna, Tenerife, Spain
\and Universit\`a di Udine and INFN Trieste, I-33100 Udine, Italy
\and National Institute for Astrophysics (INAF), I-00136 Rome, Italy
\and ETH Z\"urich, CH-8093 Z\"urich, Switzerland
\and Japanese MAGIC Group: Institute for Cosmic Ray Research (ICRR), The University of Tokyo, Kashiwa, 277-8582 Chiba, Japan
\and Technische Universit\"at Dortmund, D-44221 Dortmund, Germany
\and Croatian MAGIC Group: University of Zagreb, Faculty of Electrical Engineering and Computing (FER), 10000 Zagreb, Croatia
\and IPARCOS Institute and EMFTEL Department, Universidad Complutense de Madrid, E-28040 Madrid, Spain
\and Centro Brasileiro de Pesquisas F\'isicas (CBPF), 22290-180 URCA, Rio de Janeiro (RJ), Brazil
% ...
}


%
\abstract{We report the detection of pulsed gamma-ray emission from the Geminga pulsar (PSR J0633+1746) between $15\,\unit{GeV}$ and $75\,\unit{GeV}$. This is the first time a middle-aged pulsar has been detected up to these energies. 
Observations were carried out with the MAGIC telescopes between 2017 and 2019 using the low-energy threshold Sum-Trigger-II system. After quality selection cuts, $\sim 80\,\unit{hours}$ of observational data were used for this analysis. To compare with the emission at lower energies below the sensitivity range of MAGIC, $11$ years of \Fermi data above $100\,\unit{MeV}$ were also analysed. 
From the two pulses per rotation seen by \textit{Fermi}-LAT, only the second one, P2, is detected in the MAGIC energy range, with a significance of $6.3\,\unit{\sigma}$. 
The spectrum measured by MAGIC is well-represented by a simple power law of spectral index $\Gamma= 5.62\pm0.54$, which smoothly extends the \Fermi spectrum.
A joint fit to MAGIC and \Fermi data rules out the existence of a sub-exponential cut-off in the combined energy range at the $3.6\,\unit{\sigma}$ significance level.
The power-law tail emission detected by MAGIC is interpreted as the transition from curvature radiation to Inverse Compton Scattering of particles accelerated in the northern outer gap.
}


\maketitle

%
\section{Introduction}

Geminga (PSR J0633+1746) is an archetype of the radio-quiet gamma-ray pulsar population \citep{Bignami96,Caraveo14}. First detected by SAS-2 and COS-B \citep{sas1975,cos1977,Bignami92} as a bright gamma-ray source with no counterpart at any other wavelength and subsequently associated with an X-ray source \citep{bignami1983_1}, it was ultimately identified as a pulsar by ROSAT and EGRET \citep{rosat1992,egret1992}. It has a period of $P\simeq237\,\unit{ms}$ and a characteristic age of $\sim300\,\unit{ky}$. 
Two independent measurements of the distance reported $157_{-34}^{+59}\,\unit{pc}$ \citep{Caraveo96} and  $250_{-62}^{+120}\,\unit{pc}$ \citep{Faherty:2007:ApSS}, respectively. This makes Geminga one of the closest known pulsars.

The \Fermi detector measured the pulsed gamma-ray spectrum of Geminga using one year of data and found that it can be described by a power law with an exponential cut-off at $2.46\pm\,0.04\,\unit{GeV}$ \citep{fermi2010}. 
The increase of \Fermi statistics in the following years favoured a softer sub-exponential cut-off \citep{Abdo:2013:ApJS,magic2016}. 
Subsequent ground-based observations by the Imaging Atmospheric Cherenkov Telescopes (IACT) VERITAS \citep{veritas2015} and MAGIC \citep{magic2016} could not detect any significant emission above $100\,\unit{GeV}$ and $50\,\unit{GeV}$, respectively.
A $\sim 2\,\unit{deg}$ steady halo around Geminga was first detected by the MILAGRO experiment \citep{milagro2009}, and later reported by the HAWC \citep{hawk2017} and \Fermi \citep{fermiGemingaHalo} collaborations at energies above $5\,\unit{TeV}$ and $8\,\unit{GeV}$, respectively.

In this paper, we report the detection of pulsed gamma-ray emission from the Geminga pulsar by the MAGIC telescopes. 
This makes Geminga the first middle-aged pulsar detected by IACTs and the third pulsar detected by these type of telescopes after the Crab \citep{magiccrab2008} and Vela \citep{hessvela2018}. 
This detection had become possible thanks to the use of the new low-energy trigger system, dubbed Sum-Trigger-II \citep{Sumtrigger,jeza2014},
designed to improve the performance of the telescopes in the sub-$100\,\unit{GeV}$ energy range. 

In Section \ref{sec:analysis} we present the MAGIC observations and the technical innovations that were imperative for this detection. The analysis of \Fermi data is described in Section \ref{sec:fermi}. The resulting MAGIC and \Fermi light curves and spectra are presented in Section \ref{sec:results} and discussed in Section \ref{sec:discussion}. Finally, we compare our observations with the predictions of the pulsar outer gap model \citep{Cheng1986a,Romani1995} applied to Geminga in Section \ref{sec:theory}.

%
\section{MAGIC observations and data analysis}
\label{sec:analysis}

The MAGIC telescopes are two imaging atmospheric Cherenkov telescopes (IACTs) located on the Canary Island of La Palma (Spain) \citep{performance2016_1}.  
Observations of Geminga with the MAGIC Sum-Trigger-II system began in January 2017 and lasted until March 2019. Aiming for the lowest possible energy threshold, the observation zenith angle of the source was limited to below $25\,\unit{deg}$.
Data taken in the period between December~2017 and March~2018 were affected by non-optimal weather conditions and were discarded. After this selection, a total of $80\,\unit{h}$ of good quality data were available.
The observations were made in the so-called `wobble mode' \citep{fomin1994}, in which 
the telescopes were pointed at sky positions around Geminga with an offset of $0.4\,\unit{deg}$. 
Together with each event image, we recorded the event arrival time with a GPS disciplined Rubidium oscillator, which provides an absolute time stamp precision of $200\,\unit{ns}$.

The detection of the Geminga pulsar with MAGIC was possible thanks to the implementation of the Sum-Trigger-II system. The standard MAGIC trigger requires that the signals of three neighbouring camera pixels exceed a preset threshold of $\sim 4.0$ photo electrons ($phe.$). In the Sum-Trigger-II, the pixels are grouped into hexagonal-shape  cells of 19 pixels each. The analogue sum of all pixel signals within any given cell is compared against a discriminator threshold of $\sim 18$ $phe.$ Integrating the signal from a large area leads to a better signal-to-noise ratio (S/N) for very low energy showers. To counteract the effect of noise after-pulses,  which are typical for photo-multiplier tubes, the individual pixel amplitudes are clipped when exceeding $8.5$ $phe.$
The trigger geometry, thresholds, and clipping values were optimised by Monte Carlo (MC) simulations with the aim of minimizing the trigger threshold \citep{Sumtrigger}. 
Compared to the standard trigger, the trigger energy threshold of the Sum-Trigger-II is about $50\%$ lower. 
For a spectral index of $-5$,  which is similar to that of Geminga as reported here, the peak of the gamma energy distribution (threshold) is approximately $15\,\unit{GeV}$.

The MAGIC data were processed with the Magic Standard Analysis Software \citep[MARS,][]{mars2013}. To improve the analysis performance close to the Sum-Trigger-II energy threshold, 
we developed a new algorithm in which the calibration and the image cleaning are performed in an iterative procedure. 
The rest of the higher level analysis followed the standard pipeline described in \cite{performance2016_2}. This comprises the reconstruction of the energy and direction of the incoming gamma rays and the suppression of the hadronic background. Boosted decision trees and look-up tables were built for these purposes, using gamma-ray simulated shower events following the trajectory of Geminga in the sky and  background events from dedicated observations.

For the timing analysis, the pulsar rotational phases of the events were computed using the \textit{Tempo2} package \citep{tempo2}. 
An ephemeris for Geminga covering the MAGIC observations was obtained from the analysis of \Fermi data \citep{Kerr2015}.

%
\section{\Fermi data and analysis}
\label{sec:fermi}

To characterise the Geminga emission at energies lower than those accessible to MAGIC, we analysed $10.6$ years (from MJD $54682$ to $58569$) of public \Fermi data across the  energy range from $100\,\unit{MeV}$ to $2\,\unit{TeV}$. We processed this data set using the P8R2\_SOURCE\_V6 instrument response functions and the {\it Fermi} Science Tools version v11r5p3.
Events were selected within a circular region of interest (ROI) of $15^{\circ}$ centred at the
pulsar position (R.A.=$06^h 33^m 54.29^s$, Dec=$17^\circ 46^{'} 14.88^{''}$). 
We selected `Source' class events that were recorded only when the telescope was in nominal science mode. 
The pulsar rotational phase and barycentric corrections of the events were computed with {\it Tempo2}, using the same ephemeris as for the MAGIC data analysis.
The pulsar light curve was produced applying an additional energy dependent angular cut, according to the approximation of the \Fermi Pass8 point spread function for a 68\% confinement radius \citep{3FGL}. 

For the spectral reconstruction, a binned likelihood analysis was performed making use of the {\it pyLikelihood} python module of the {\it Fermi} Science Tools. Each of the two emission peaks of the Geminga light curve, P1 and P2, were analysed separately. We started the likelihood fits by including all sources in the ROI from the third {\it Fermi} Source Catalogue \citep{3FGL} in the spectral-spatial model. The spectral parameters for sources with a significance higher than $5\,\unit{\sigma}$ and located within 5 deg of the centre of the ROI were left free. Also, we let the normalisation factor of the isotropic and Galactic background models free. For the rest of the sources, all parameters were fixed to their catalogue values. In a second step, all sources with $TS<4$ were removed from the model. For the calculation of the spectral points, we repeated the procedure in each energy bin using a power law with the normalisation factor free and the spectral index fixed to $2$.

%
\section{Results}
\label{sec:results}

\subsection{Light curves}

\begin{figure*}
    \centering
    \includegraphics[width=\linewidth]{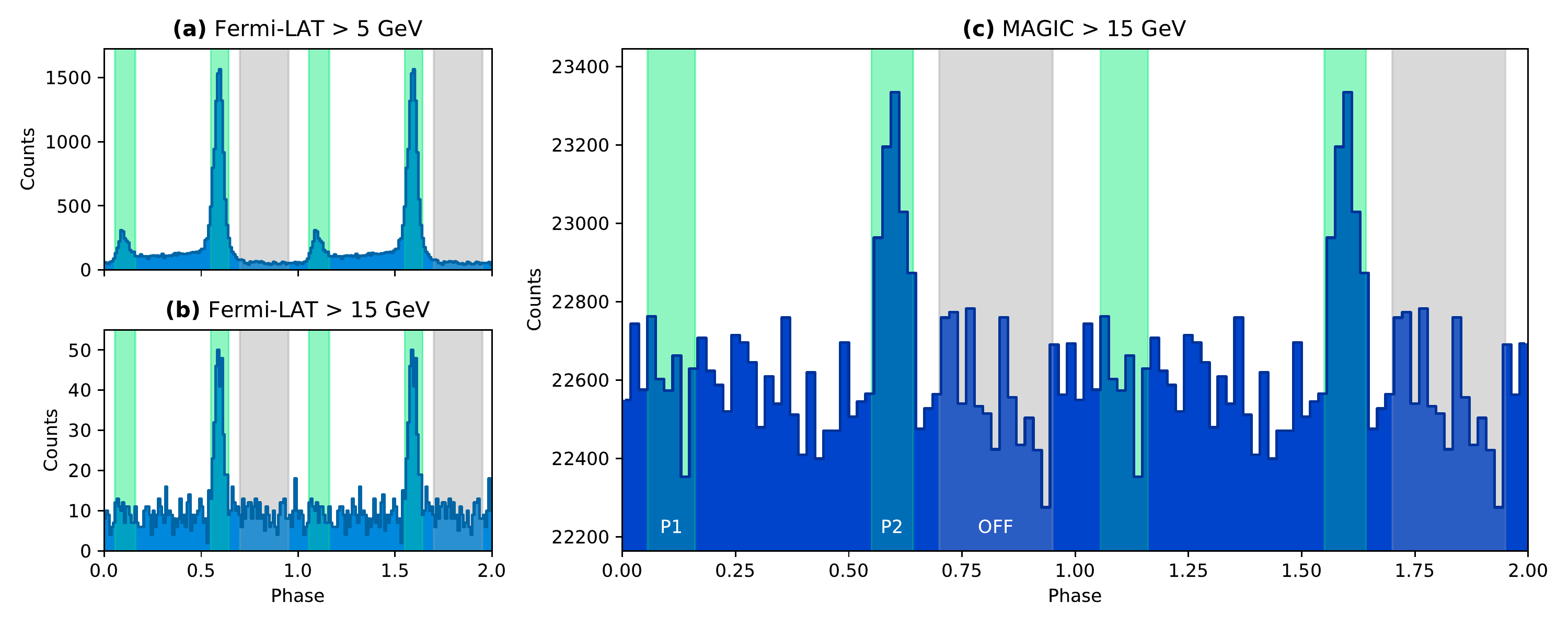}
    \caption{
    Geminga light curves measured by \Fermi (panels {\bf a} and {\bf b}), and by MAGIC (panel {\bf c}). For clarity, two rotation cycles are shown. 
    The green-shaded regions highlight the phase intervals corresponding to the $P1$ and $P2$ emission, obtained from the fits to \Fermi data above $5\,\unit{GeV}$ and $15\,\unit{GeV}$, respectively. The phase region from which the background is estimated is shown by the grey band.
    These signal regions were later applied to the analysis of MAGIC data. P2 is detected with MAGIC at a significance level of $6.25\,\unit{\sigma}$. No significant signal is detected from P1 in the MAGIC energy range.
    }
    \label{fig:phaseogram}
\end{figure*}

The light curves shown in Fig. \ref{fig:phaseogram} are produced by phase folding \Fermi photons and MAGIC events using the same pulsar ephemeris. The two well-known Geminga emission peaks, $P1$ and $P2$, are clearly visible above $5\,\unit{GeV}$ in \Fermi data. At higher energies, only P2 is detected by \Fermi, which is in agreement with the high-energy light curves shown in \cite{FermiCatalog10GeV}.
To characterise each peak at energies as close as possible to the MAGIC energy range, we fit them to symmetric Gaussian functions, using \Fermi events above $5\,\unit{GeV}$ for P1 and above $15\,\unit{GeV}$ for P2. The corresponding light curves are shown in Fig. \ref{fig:phaseogram}, panels $(a)$ and $(b)$. The phase signal regions for the analysis of MAGIC data are then defined as the $\pm 2\,\unit{\sigma}$ intervals around the fitted peak positions. 
For estimating the background, we considered the off-pulse region between P2 and P1,
where no emission is expected from the pulsar, starting $6\,\unit{\sigma}$ away from each peak's centre. Table \ref{tab:regions} summarises the signal and background regions used for the MAGIC analysis. 

The MAGIC light curve for events with reconstructed energies above $15\,\unit{GeV}$ is shown in Fig. \ref{fig:phaseogram}, panel $(c)$. It was obtained after applying energy-dependent gamma and hadron separation cuts, trained on MC simulated gamma-ray showers. The number of excess events for each emission peak and the corresponding significances were computed using Eq. $17$ in \cite{lima1983} and they are tabulated in Table \ref{tab:regions}. 
Emission from P2 is detected with MAGIC at a significance level of $6.25\,\unit{\sigma}$, corresponding to $2457$ excess events over a scaled background of $112018$ events. A region-independent signal test with the $H$-test and $Z^2_{10}$-test \citep{dejager2010} results in $4.8\,\unit{\sigma}$ and $5.2\,\unit{\sigma}$ significances, respectively. The analysis of MAGIC events in the phase region of P1 does not reveal any significant signal in this energy range.  

\begin{table}
\centering
\begin{tabular}{rlrr}
\hline\hline
\multicolumn{2}{c}{\bf Phase region} & \bf Excess & \boldmath \bf $\sigma$\\ 
\hline
\emph{P1}  & (0.056~\textendash~0.161) &  116.7 & 0.27\\
\emph{P2}  & (0.550~\textendash~0.642) & 2457.3 & 6.25\\
\emph{OFF} & (0.700~\textendash~0.950) &  ~      & ~ \\
\hline
\end{tabular}
\caption{Definition of the signal ($P1$, $P2$) and background ($OFF$) regions derived from the analysis of the \Fermi light curves shown in Fig. \ref{fig:phaseogram}. The last two columns refer to the number of excess events and the significance $\sigma$ (following the Li\&Ma definition) obtained in the analysis of MAGIC data.}
\label{tab:regions}
\end{table}

\subsection{Energy spectrum}
 
\begin{figure*}
    \centering
    \includegraphics[width=0.9\linewidth]{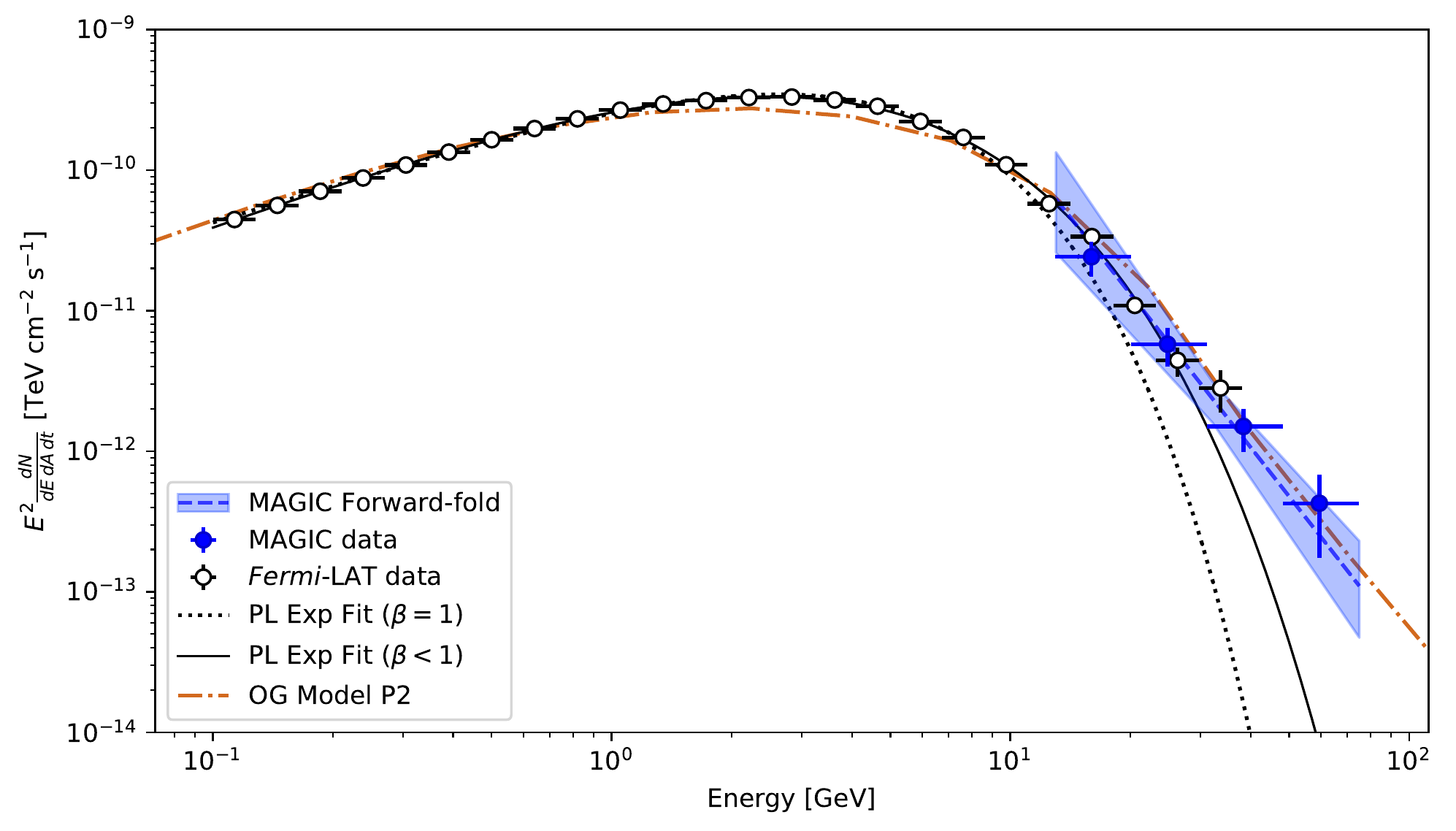}
    \caption{SED of the second emission peak, P2, of the Geminga pulsar  measured by the MAGIC telescopes (filled circles) and the \Fermi (open circles). The MAGIC spectral points were calculated by unfolding the reconstructed energy spectrum by means of the Tikhonov regularisation method. Dashed blue line shows the forward folding fit to the MAGIC data assuming a power law and the blue-shaded area represents the $1\,\unit{\sigma}$ confidence interval around the power-law fit. 
    Dotted and continuous black lines represent the combined fit to MAGIC and \Fermi data to a power law with an exponential or sub-exponential cutoff, respectively.  
    Dotted-dashed orange line shows the P2 spectrum predicted in the stationary outer gap model applied to Geminga for a magnetic dipole moment $\mu= 1.4\,\mu_{\rm d}$, an angle between the magnetic and the rotational axis of the star, $\alpha= 30^\circ$, and an observer's viewing angle, $\zeta= 95^\circ$. 
    }
    \label{fig:spectrum}
\end{figure*}

Figure \ref{fig:spectrum} shows the Spectral Energy Distribution (SED) of Geminga for P2, obtained after the analysis of \Fermi (open circles) and MAGIC data (filled circles). 
The spill-over effect due to the soft spectral index has been carefully taken into account by unfolding the MAGIC energy spectrum using the Tikhonov regularisation method  \citep{MAGIC_Unfolding_2007}, and cross-checked with a forward-folding procedure. 
The resulting MAGIC unfolded spectral points (filled circles in Fig. \ref{fig:spectrum}) are reported in Table \ref{tab:magicpoints}. 
The dashed blue line shows the forward-folding power-law fit, $F_0(E/E_0)^{-\Gamma}$, performed on the distribution of MAGIC excess events. 
The blue butterfly represent the $1\,\unit{\sigma}$ statistical uncertainty confidence interval of the fit. 
The spectrum measured by MAGIC in the energy range $15 - 75\,\unit{GeV}$ is well-represented by the power law, with an associated $\chi^2=15.27$ with $15$ degrees of freedom. 
The obtained spectrum is in agreement with the upper limits previously reported by MAGIC \citep{magic2016}. 
The resulting fit parameters are reported in Table \ref{tab:fits}. 
The spectral index $\Gamma=5.62\pm0.54$ (statistical errors only) is the softest ever measured by MAGIC from any source. 

\begin{table}
\centering
\begin{tabular}{cccc}
\hline\hline
\boldmath $E_{low}$ & \boldmath $E$ & \boldmath $E_{hi}$ & \boldmath $E^2dN/(dE\,dA\,dt)$ \\ \hline
$12.9$ & $16.0$ & $20.1$  &  $(2.42 \pm 0.67)\cdot10^{-11}$\\
$20.1$ & $24.8$ & $31.1$  &  $(5.77 \pm 1.79)\cdot10^{-12}$\\
$31.1$ & $38.4$ & $48.2$  &  $(1.50 \pm 0.51)\cdot10^{-12}$\\
$48.2$ & $59.5$ & $74.8$  &  $(4.30 \pm 2.50)\cdot10^{-13}$\\
\hline
\end{tabular}
\caption{MAGIC SED points. The energy bin edges ($E_{low}$, $E_{hi}$), as well as their centre position, $E$, are in units of $\unit{GeV}$. SED values are in $\unit{TeV\,cm^{-2}\,s^{-1}}$. }
\label{tab:magicpoints}
\end{table}

%
\section{Discussion}
\label{sec:discussion}

We performed a joint fit of MAGIC and \Fermi spectral points in the combined energy range, from $100\,\unit{MeV}$ to $75\,\unit{GeV}$, by using a power law with an exponential cut-off function: 
\begin{equation}
\label{eqn:expocut}
F(E)=F_0 \left(\nicefrac{E}{E_0}\right)^{-\Gamma} \exp{\left(-(\nicefrac{E}{E_c})^{\beta}\right)}
,\end{equation}
where $E_0$ is the energy scale, $\Gamma$ the spectral index, $E_c$ the cut-off energy, and $\beta$ the cut-off strength. We considered two different cases: a pure exponential cut-off, $\beta = 1$, and the general case in which the $\beta$ parameter is set free. The parameters resulting from the fits are given in Table \ref{tab:fits}.
The best fit is found for $\beta=0.738\pm0.013$.
In a likelihood-ratio test versus the free exponential cut-off model, the pure exponential case can be rejected with ${\rm TS} = -2\Delta\log\mathcal{L}=336.6$, 
which, according to a chi-square distribution with one degree of freedom, corresponds to a significance of $18.3\,\unit{\sigma}$. 
Also, the sub-exponential cut-off is disfavoured by the data at the level of $3.6\,\unit{\sigma}$, according to a goodness-of-fit chi-square test.

In order to assess whether there is any preference for curvature in the
high energy tail of the spectrum, we fitted MAGIC and \Fermi spectral points above $10\,\unit{GeV}$ to a log-parabola model, $F_0(E/E_0)^{-\Gamma-\beta\log(E/E_0)}$, obtaining a best-fit value for the curvature index of $\beta\,=\,0.99_{-0.84}^{+0.67}$. 
We performed a likelihood-ratio test between this model and a power law. This results in ${\rm TS}=1.7$ with 1 degree of freedom, corresponding to $1.3\,\unit{\sigma}$ against the power-law model. 
This shows that the log-parabola model is not significantly preferred over the power-law one. The power-law index derived from the joint fit, $\Gamma\,=\,5.18\pm0.15$, is compatible with the one obtained with MAGIC data alone.

The effect of systematic uncertainties on the MAGIC spectral reconstruction has been studied. 
A $5\%$ change in the estimated energy of the events has little effect in the reconstructed spectral index (below $1\%$), but makes the MAGIC fluxes fluctuate by up to $20\%$.
This results from the combined effect of the softness of the Geminga spectrum and the steeply falling effective collection area close to the MAGIC energy threshold. 
The joint fit with \Fermi data in the overlapping energy range helps to constrain these uncertainties. To test this we introduced a MAGIC flux scale factor, s, as a nuisance parameter in the fits. Maximum-likelihood values of the scale parameter are always found to be compatible with unity within the uncertainties, with $s=0.88\pm0.11$ and $s=0.98\pm0.14$ for the sub-exponential and the log-parabolic fits, respectively. Likelihood-ratio tests versus a model with no scaling provide, in both cases,  ${\rm TS}\,<\,1.0$. We conclude that the energy calibration of MAGIC with respect to \Fermi is accurate. 
 
\begin{table*}
\centering
\begin{tabular}{cccccccc}
\hline\hline
 \boldmath Data & \boldmath Function & \boldmath $F_0$ & \boldmath $E_0$ & \boldmath $\Gamma$  & \boldmath $E_c$ & \boldmath $\beta$ & \boldmath $-2\log\mathcal{L}$ \\
 \hline
\rule{0pt}{2.5ex} MAGIC & PL & $(2.28\pm0.74)\cdot10^{-9}$ & $32.15$ & $5.62\pm0.54$ & - & - & -\\
F \& M & PL Exp & $(0.357\pm0.002)\cdot10^{-3}$ & $1.00$ & $1.089\pm0.003$ & $2.88\pm0.02$ & 1 &  $388.2$  \\
F \& M & PL Subexp & $(0.552\pm0.019)\cdot10^{-3}$ & $1.00$ & $0.910\pm0.013$ & $1.44\pm0.08$ & $0.738\pm 0.013$ &  $51.6$ \\
F{>10} \& M   & PL & $(0.29 \pm 0.04)\cdot10^{-8}$ & 32.15 & $5.18\pm0.15$ & - &- & $11.1$ \\
F{>10} \& M   & LP & $(0.21 \pm 0.06)\cdot10^{-8}$ & 32.15 & $6.4^{+1.1}_{-0.9}$ & - & $0.99_{-0.84}^{+0.67}$ & 9.4 \rule[-1.2ex]{0pt}{0pt} \\
\hline
\end{tabular}
\caption{Results from the spectral fits performed to the MAGIC data alone (first row) to a power law function (PL), and from the joint fits to \Fermi and MAGIC data (abbreviated as F \& M) to power laws with exponential (PL Exp) or sub-exponential (PL Subexp) cut-offs, and to a log-parabola model (LP). 
For the last two fits only \Fermi spectral points above $10\,\unit{GeV}$ were used. The normalisation factor $F_0$ is given in units of $\unit{TeV^{-1}\,cm^{-2}\,s^{-1}}$. The normalisation energy, $E_0$, and the cut-off energy, $E_c$, are given in units of $\unit{GeV}$. Also, $\Gamma$ refers to the PL spectral index, and $\beta$ to the cut-off strength, except for the LP case for which $\beta$ represents the curvature of LP.
The quoted uncertainties are statistical at a $1\sigma$ confidence level.}
\label{tab:fits}
\end{table*}

%
\section{Modelling the high-energy emission}
\label{sec:theory}

Two main radiation processes are considered to be responsible for the gamma-ray emission detected in pulsars: synchro-curvature radiation or Inverse Compton Scattering (ICS), or a combination of both. The first can explain the exponential cut-offs at a few GeV seen in the vast majority of \Fermi pulsars, while the second process may account for the power-law spectral tail detected in the Crab pulsar up to TeV energies \citep{MagicCrabTeV:2016:AA}. 

We compare our observational results with the predictions of the stationary three-dimensional  pulsar outer gap (OG) model \citep{Hirotani:2006:ApJ,Hirotani:2013:ApJ}.
We assume that the magnetic field lines are given by the rotating vacuum dipole solution \citep{Cheng:2000:ApJ, veritas2015}, and we solve  Gauss's law, the stationary Boltzmann equations for electrons and positrons, and the radiative transfer equation of the emitted photons from IR to very-high-energy (VHE) gamma rays. 
Accordingly, we can obtain the pulse profile and the phase-resolved spectrum of the emitted photons, by setting the following five parameters of the neutron star (NS): the rotational angular frequency, $\Omega_{\rm F}$; the surface temperature, $T$;  the area of the star, $A$; the magnetic dipole moment, $\mu$; and the angle $\alpha$ between the NS magnetic and rotational axes. 
The NS rotational period, $P$, is an observable and readily gives $\Omega_{\rm F}= 2\pi / P$.
Using the soft X-ray data \citep{Halpern:1997:ApJ}, we constrain the NS surface emission with temperature $kT= 49.74\,\unit{eV}$ and area $A=0.5085 \times 4\pi r_0{}^2$, where $4\pi r_0{}^2$ denotes the whole NS surface area measured by a distant observer. The distance to source is assumed to be $250\,\unit{pc}$ \citep{Faherty:2007:ApSS}.
We also include in the seed X-ray spectrum a harder component ($kT\sim185\,\unit{eV}$) associated with the heated polar cap region discussed in \cite{caraveo2004}.
The value of $\mu$ will not be very different from its dipole value $\mu_{\rm d}$, which is constrained by $P$ and its temporal derivative, $\dot{P}$, under the assumption of magnetic dipole radiation. Therefore, $\mu / \mu_{\rm d}$, $\alpha$, and the observer's viewing angle with respect to the rotation axis, $\zeta$, remain as free parameters. We constrain these parameters by comparing the predicted pulse profile and phase-resolved spectrum with the MAGIC and \Fermi observations.

For the Geminga pulsar, unlike in young pulsars like the Crab, ICS is negligible in the outer magnetosphere because the IR photon fields are too weak. As a result, the positrons accelerated outward in the gap, produces negligible VHE fluxes. Nevertheless, gap-accelerated electrons continue propagating towards the star to efficiently up-scatter soft X-ray photons from the  NS surface by head-on collisions. Accordingly, if we are observing Geminga nearly perpendicularly to its rotation axis, the inward VHE photons emitted by the electrons would appear in the same rotational phase as the outward HE photons emitted by the positrons. 

To explain the MAGIC flux and the double-peaked pulse profile observed at lower energies by \Fermi with a peak separation of $183^\circ$ \citep{Abdo:2013:ApJS}, we find that $\mu= 1.4$\,$\mu_{\rm d}$, $\alpha \sim 30^\circ$ and $95^\circ < \zeta < 100^\circ$ are necessary. 
A similar viewing angle of $\zeta = 90^\circ$ was also obtained in \cite{Pierbattista15}, by fitting the \Fermi light curve on the basis of the OG model.
The  dotted-dashed orange line in Fig. \ref{fig:spectrum} shows the predicted Geminga flux from the P2 phase region defined in Table \ref{tab:regions} for a viewing angle $\zeta= 95^\circ$. 
The OG solutions predicting emission in the MAGIC energy range tend to under predict the \Fermi flux at few GeV. This may indicate the limitation of stationary OG models, suggesting the need for non-stationary particle-in-cell simulations of pulsar magnetospheres\footnote{Particle-in-cell simulations (see e.g. \cite{Brambilla:2018:ApJ} and references therein), are currently limited to much weaker magnetic field strengths than the actual values found in pulsar magnetospheres. Consequently, they cannot be used at present to probe pulsar emission above GeV energies.}. 
At the viewing angle of $\zeta=95^\circ$, below $40\,\unit{GeV}$, the flux of P2 is dominated by the outward photons emitted in the northern OG via mainly the curvature process. Above $40\,\unit{GeV}$, the P2 flux is instead dominated by the inward photons emitted via ICS in the same northern OG.
The emission from the southern OG is relatively small for P2, whereas for P1, the role of northern and southern OGs are exchanged. 
At $\zeta=100^\circ$ the inward ICS emission dominates the outward one above $100\,\unit{GeV}$. At $\zeta \ge 105^\circ$, the ICS component becomes negligible. 
It also follows that the present stationary OG model predicts an extension of the ICS pulsed component above the energies reported by MAGIC.

%
\section{Summary} 
In this paper, we present the detection of pulsed gamma-ray emission from the Geminga pulsar with the MAGIC telescopes. The emission coincides in pulse phase with the position of $P2$ and is detected up to $75\,\unit{GeV}$. This makes Geminga the first middle-aged pulsar detected up to such energies. 
The spectrum measured by MAGIC is well-described by a power law of spectral index $\Gamma= 5.62\pm0.54$.  
A joint fit to MAGIC and \Fermi data rules out the existence of an exponential cut-off in the combined energy range. A sub-exponential cut-off is also disfavoured at the $3.6\,\unit{\sigma}$ level. 
According to the outer gap model, the emission detected by MAGIC implies that we are observing Geminga nearly perpendicularly to its rotation axis and that the emission originates in the northern outer gap. The energy range covered by MAGIC would correspond to the transition from curvature radiation by outward accelerated positrons 
to ICS by electrons accelerated 
towards the star. The ICS component is predicted to extend above the energies detected by MAGIC. 
This should be confirmed by future Geminga observations by IACTs, as well as by non-stationary pulsar gap models.

%
\begin{acknowledgements}
We would also like to thank the Instituto de Astrof\'{\i}sica de Canarias for the excellent working conditions at the Observatorio del Roque de los Muchachos in La Palma. The financial support of the German BMBF and MPG; the Italian INFN and INAF; the Swiss National Fund SNF; the ERDF under the Spanish MINECO (FPA2017-87859-P, FPA2017-85668-P, FPA2017-82729-C6-2-R, FPA2017-82729-C6-6-R, FPA2017-82729-C6-5-R, AYA2015-71042-P, AYA2016-76012-C3-1-P, ESP2017-87055-C2-2-P, FPA2017-90566-REDC); the Indian Department of Atomic Energy; the Japanese ICRR, the University of Tokyo, JSPS, and MEXT;  the Bulgarian Ministry of Education and Science, National RI Roadmap Project DO1-268/16.12.2019 and the Academy of Finland grant nr. 320045 is gratefully acknowledged. This work was also supported by the Spanish Centro de Excelencia ``Severo Ochoa'' SEV-2016-0588 and SEV-2015-0548, the Unidad de Excelencia ``Mar\'{\i}a de Maeztu'' MDM-2014-0369 and the ``la Caixa'' Foundation (fellowship LCF/BQ/PI18/11630012), by the Croatian Science Foundation (HrZZ) Project IP-2016-06-9782 and the University of Rijeka Project 13.12.1.3.02, by the DFG Collaborative Research Centers SFB823/C4 and SFB876/C3, the Polish National Research Centre grant UMO-2016/22/M/ST9/00382, by the Brazilian MCTIC, CNPq and FAPERJ, and at HKU by a GRF grant (Project 17307618) from the Hong Kong Government.
\end{acknowledgements}

%

\end{document}